\documentstyle[12pt]{article}

\begin{document}
\renewcommand{\thefootnote}{\fnsymbol{footnote}}

\def\lsim{\mathrel{\raise.3ex\hbox{$<$\kern-.75em\lower1ex\hbox{$\sim$}}}}
\def\gsim{\mathrel{\raise.3ex\hbox{$>$\kern-.75em\lower1ex\hbox{$\sim$}}}}

\begin{flushright}
HUE--99/2 \\
February~~~~2000  \\

\end{flushright}

\vspace{1cm}
\begin{center}
{\LARGE\bf Anomalous U(1) Symmetry and Sparticles} \\
\vspace{3cm}
Shun'ichi Tanaka \\
\vspace{0.8cm}
\it{Department of Physics,} \\
\it{Hyogo University of Education, Yashiro-cho, Hyogo 673-1494, Japan} \\
\vspace{3cm}

\begin{abstract}
  Using anomalous $U(1)$ symmetry the quark mass texture is
determined uniquely.  We analyze squark mass 
spectrum  based on the above mass matrices and discuss the possibility
 to solve the problems of FCNC
and CP caused by complex phases of soft terms,  including the viewpoint
 of M theory.
\end{abstract}
\end{center}
\noindent
\hspace*{1cm} PACS number(s): 11.30.Hv, 14.80.Dq, 14.80.Ly.\\
\hspace*{1cm} Keyword(s): anomalous U(1) symmetry, FCNC, squark mass. 
\clearpage
 
{\bf Introduction}\par
The minimal supersymmetric standard model (MSSM) does not contradict
 any experiment till now.  However, it does not explain the origin of mass
 hierarchy and the problem of flavor-changing neutral current, the CP
violation problem in soft terms, the origin of baryogenesis and so on.
  In string theory, which often has anomalous $U(1)$ gauge symmetry,
 there have been many attempts to explain the above problems  by
 using this symmetry \cite{R1}.   We investigate these subjects 
 including a viewpoint of M-theory which has developed very rapidly.
In the Ho\v{r}ava-Witten construction \cite{H1}, 11-dimensional supergravity
 (low-energy
 limit of M-theory) is compactified on $M_{4}\times S^{1}/Z_{2}\times X$,
where $X$ is a 6-dimensional Calabi-Yau manifold, $M_{4}$  is a 4-dimensional
Minkowski space, and the fifth dimension is compactified on a line segment
($S^{1}/Z_{2}$)  whose length, $\pi\rho$, is larger than the ``radius'' of
 the Calabi-Yau volume.  In this picture, the observable and hidden gauge 
degrees of freedom (the former is the surviving subgroup of $E_{8}$, and
 the latter is $E^{\prime}_{8}$ or its subgroup) live on two distinct
 4-dimensional ``walls'', a distance $\pi\rho$ apart.  Standard model gauge
 fields and charged matter are confined to one wall (or it may be on a 
``non-perturbative'' five-brane).  Gravity lives on the 5-dimensional
 ``bulk''.  Distinctive features to take account of 
M-theory are that the string coupling is strong and because there is
 a separating bulk
between the observable and hidden sectors, there live vector- and hyper-
multiplets.  \par
  Previously we examined mass
 spectrum of quarks and charged leptons \cite{K1} based on $Z_{3}$
 orbifold model using flipped $SO(10)$ symmetry because in this model
 flipped $SO(10)$ exists uniquely as a group of GUT.  The mass spectrum
 could be reproduced,  but simpler is to use anomalous $U(1)$ symmetry that
 acts as a horizontal symmetry \cite{R2}.  Although various anomalous
 $U(1)_{X}$
 charges of quarks have been assigned to
 fit the low-energy mass spectrum using the renormalization group method, 
they were  rather devoid of definiteness.  However, as pointed out in
\cite{E1}, $U(1)_{X}$ charges can be determined almost uniquely from
the CKM angles.  In \cite{E1} this $U(1)_{X}$ symmetry is embedded
into a larger gauge group (GUT).  We treat the anomalous $U(1)_{X}$ 
symmetry separately here and apply the result to the sector of
superpartners\footnote{When we embed the anomalous $U(1)_{X}$
group in a GUT group, the problem of the existence of an
 adjoint Higgs boson as well as complexity to include various representations
 of Higgs bosons and how to break the symmetry is difficult to solve.
  Here we consider only $U(1)_{X}$ group ``simple-mindedly.''}.
Because the $U(1)_{X}$ symmetry is not observed 
at low energies, it must be broken.  The Higgs mechanism is used
 to break the $U(1)_{X}$ gauge symmetry.  We denote this kind of Higgs
 field $\theta$.  The electroweak singlet fields $\theta$
 may appear in vectorlike pairs or as chiral individuals.
If they appear in vectorlike pairs, they would obtain in general very
 large mass of order the Planck mass.
We take $\theta$ to be a chiral superfield in this paper.
  Then the mixed anomalies of
 the $U(1)_{X}$ symmetry are necessarily nonzero and must be cancelled using
 the Green-Schwarz mechanism \cite{G1}.  This incidentally fixes the weak
 mixing angle
 without recourse to GUT \cite{I1}.  Moreover, the value of the parameter
 $\varepsilon \equiv \langle \theta \rangle / M_{P}$, where $M_{P}$ is
the four-dimensional reduced Planck mass, is determined definitely from
 the $D$-term.
The structure (texture) of mass matrices will change little if two or more
 chiral fields $\theta$ are included.  So we consider a single chiral
 field in this paper.  We denote the $U(1)_{X}$ chiral charge of $\theta$, 
 $X(\theta)=-q_{\theta}<0$.
$U(1)_{X}$ charges of matter fields are fractional in general.
We assume chiral flavor charges of matter superfields
 $\phi_{i}$ are greater than or equal to zero, $X(\phi_{i})\geq 0$ 
\footnote{Of course, the important thing here is the form of the
 superpotential and not the $U(1)_{X}$ charge. However,when considering
the mass of sparticles and not to break the symmetries of the standard model
at high energy,this assumption is necessary.} .  \par 
In most of the previous papers, in which the anomalous $U(1)$ symmetry was
 used, the 
mass term of $\theta$ was inserted explicitly and the mass scale was set at
the electroweak scale not necessarily definite, and
 furthermore, in many cases supersymmetry is broken by the $D$-term using that
mass term \cite{D1}.  
However, taking supergravity into consideration at the same time, the
 situation improves.  We consider here that supersymmetry is broken 
locally by the gaugino condensation in the hidden sector.  Then $\theta$,
 squarks and gauginos in the observable sector get mass of the order of the
 gravitino mass.  The $D$-term becomes nonzero concurrently with it, and 
first two generations of squarks will get much larger mass than the gravitino 
mass because the value of the $D$-term is greater than the gravitino mass.\par

  In what follows we first define the form
of the $U(1)_{X}$ charge.  Next we determine the mass
 matrices of up- and down-quarks from the mass of the quarks and
the CKM matrix.  However, there remains a few versions at this stage. 
 By considering the Higgs sector we can fix the charges definitely. 
 These charges affect the mass
 spectrum of sparticles.  We derive the values of masses of sparticles and
 discuss that
this scheme will avoid the problems of FCNC  \cite{C1} and CP breaking by
soft terms \cite{C1},\cite{N1}, and suggest a model
 to solve the problem of baryogenesis.  \par 

\bigskip
{\bf$U(1)_{X}$ charge}\par
The superpotential for the up quarks is given by
\begin{equation}
W\sim Q_{i}u^{c}_{j}H_{2} \left(\frac{\theta}{M_{P}}\right)
 ^{n^{(u)}_{ij}},
\end{equation}
where $i$ represents the generation.  We neglect numerical factors because
in this approach we are interested in the order of  magnitude.  From
charge conservation one obtains
\begin{equation}
   q_{\theta}n^{(u)}_{ij}=X(Q_{i})+X(u^{c}_{j})+X(H_{2}) .  \label{n33}
\end{equation}
As for top mass, there is an infrared stable quasi-fixed point as long as
the Yukawa coupling of the top is of order one \cite{F1}.  The predicted
 value fits experimental data very well.  So we assume  $ n^{(u)}_{33}=0$, 
 and $ n^{(u)}_{ij} \geq 0  \:\:\:  $for$ \  (i,j)\neq (3,3)$ . 
Effective Yukawa couplings are given by
\begin{equation}
Y^{(u)}_{ij}\sim\left(\frac{<\theta>}{M_{P}}\right)^{n^{(u)}_{ij}} .
\end{equation}
The $3 \times 3$ mass matrix of the up quarks are written as
\begin{equation} 
 \frac{M_{u}}{m_{t}}\sim\left(\varepsilon^{n^{(u)}_{ij}}\right). 
\end{equation}
Denoting the $U(1)_{X}$ charges of quarks and electroweak Higgs fields  as
\begin{eqnarray*}
   X(Q_{i})=\alpha_{i}, \:  X(u^{c}_{i})=\beta_{i}, \: 
   X(d^{c}_{i})=\gamma_{i}, 
\end{eqnarray*}
\begin{equation}
    X(H_{1})=h_{1}, \:   X(H_{2})=h_{2} , 
\end{equation}
Eq.(2) is written as
\begin{equation}
   n^{(u)}_{ij}=\frac{1}{q_\theta}(\alpha_{i}+\beta_{j}+h_{2}) . 
\end{equation}
If we factor out
\begin{equation}
   m_{t}=\lambda_{t}<H_{2}>\varepsilon^{n^{(u)}_{33}}=\lambda_{t}<H_{2}> 
\end{equation}
from the mass matrix, then
\begin{equation}
 (n^{(u)}_{ij})=\frac{1}{q_{\theta}}
 \left(
  \begin{array}{ccc}
 (\alpha_{1}-\alpha_{3})+(\beta_{1}-\beta_{3}) & (\alpha_{1}-\alpha_{3})+
 (\beta_{2}-\beta_{3}) & \alpha_{1}-\alpha_{3} \\
 (\alpha_{2}-\alpha_{3})+(\beta_{1}-\beta_{3}) & (\alpha_{2}-\alpha_{3})
 +(\beta_{2}-\beta_{3}) & \alpha_{2}-\alpha_{3} \\
 \beta_{1}-\beta_{3} & \beta_{2}-\beta_{3} & 0
 \end{array}
\right) 
\end{equation}
For the down quarks we obtain the same expression with $\gamma_{i}$
instead of $\beta_{i}$  except for the prefactors. \par
 Hereafter, for simplicity, we assume that $-q_{\theta}=-1$, and that
the chiral charges of matter fields are integer.\\

\bigskip
{\bf quark sector} \par
 We neglect the complex (CP) phase in the quark sector, for simplicity.
To determine the form of quark matrices, the form of the CKM matrix must be
fixed first. 
 The Wolfenstein parametrization of the CKM matrix is given by
\begin{equation}
\left(
\begin{array}{ccc}
1 & \lambda & \lambda^{3} \\
         -\lambda & 1 & \lambda^{2} \\
         \lambda^{3} & -\lambda^{2} & 1
\end{array}
\right)
\end{equation}
in terms of the Cabibbo angle $\lambda$ ($\lambda \simeq 0.22$ ), with all
prefactors of order one.  Although the (3, 1) component can be chosen
$\lambda^{4}$ instead of $\lambda^{3}$ according to experimental data, 
$\lambda^{4}$ is excluded by the argument below of charge conservation.      
If we assume the Grand Desert Scenario, namely no significant matter
 between the TeV scale and string scale around $10^{16}$GeV,
the fermion masses satisfy
\begin{equation}
  \frac{m_{u}}{m_{t}}\sim\lambda^{7-8}  ;\:
 \frac{m_{c}}{m_{t}}\sim\lambda^{4} ;\: \frac{m_{d}}{m_{b}}\sim\lambda^{4} ;\:
\frac{m_{s}}{m_{b}}\sim\lambda^{2} . 
\end{equation}
As mixing angles are small, at the scale $M_{string}$
\begin{equation}
  {\it M}^{diag}_{u}\sim
\left(
\begin{array}{ccr}
 \lambda^{8}(\lambda^{7}) & 0 & 0 \\
 0 & \lambda^{4} & 0 \\
0 & 0 & 1
 \end{array}
\right)m_{t} , 
\end{equation}

\begin{equation}
  {\it M}^{diag}_{d}\sim
\left(
\begin{array}{lcr}
\lambda^{4} & 0 & 0 \\
 0 & \lambda^{2} & 0 \\
 0 & 0 & 1
\end{array}
\right)m_{b} . 
\end{equation}
The relation $\varepsilon \sim \lambda$ should hold.  We will remark on the
appropriateness of it later.
The CKM matrix is expressed as
\begin{equation}
   V_{CKM}=V^{(L)\dagger}_{u}V^{(L)}_{d} , 
\end{equation}
where unitary matrices $V^{(L)}_{u}$ and $V^{(L)}_{d}$ are defined as
\begin{equation}
  {\it M}_{D}{\it M}^{\dagger}_{D}=V^{(L)}_{d}  
 ({\it M}^{diag}_{d})^{2}V^{(L) 
\dagger}_{d} , 
\end{equation}

\begin{equation}
   {\it M}_{U}{\it M}^{\dagger}_{U}=V^{(L)}_{u}({\it M}^{diag}_{u})^{2} 
V^{(L)\dagger}_{u} . 
\end{equation}
We want to determine first the form of $M_{U}$ and $M_{D}$ from $M^{diag}$
and $V_{CKM}$.
If the following relations
\begin{equation} 
  [{\it M}_{D}{\it M}^{\dagger}_{D},V^{(L)}_{u}]=0, \:\:
    [{\it M}_{U}{\it M}^{\dagger}_{U},V^{(L)}_{d}]=0
\end{equation}  
are satisfied, ${\it M}_{D}{\it M}^{\dagger}_{D}$  and
  ${\it M}_{U}{\it M}^{\dagger}_{U}$  are uniquely determined as follows,
\begin{equation}
   {\it M}_{U}{\it M}^{\dagger}_{U}=V^{\dagger}_{CKM}
 ({\it M}^{diag}_{u})^{2} 
   V_{CKM}=
\left(
\begin{array}{lcc}
\lambda^{6} & \lambda^{5} & \lambda^{3} \\
\lambda^{5} & \lambda^{4} & \lambda^{2} \\
\lambda^{3} & \lambda^{2} & 1
\end{array}
\right)m^{2}_{t} , 
\end{equation}

\begin{equation}
    {\it M}_{D}{\it M}^{\dagger}_{D}=V_{CKM}({\it M}^{diag}_{d})^{2} 
   V^{\dagger}_{CKM}=
\left(
\begin{array}{lcc}
\lambda^{6} & \lambda^{5} & \lambda^{3} \\
\lambda^{5} & \lambda^{4} & \lambda^{2} \\
\lambda^{3} & \lambda^{2} & 1
\end{array}
\right)m^{2}_{b} . 
\end{equation}
Or according to Elwood, Irges and Ramond \cite{E1}, if we assume that
 $V^{(L)}_{u}$
 and $V^{(L)}_{d}$  have the same form as that of $V_{CKM}$ 
 with different numerical factors, we get the same form as above, since 
\begin{equation}
{\it M}_{U}{\it M}^{\dagger}_{U}=V^{(L)}_{u}({\it M}^{diag}_{u})^{2} 
V^{(L)\dagger}_{u} \simeq V_{CKM}({\it M}^{diag}_{u})^{2} 
   V^{\dagger}_{CKM} ,
\label{eq:u-mass2} 
\end{equation}

\begin{equation}
 {\it M}_{D}{\it M}^{\dagger}_{D}=V^{(L)}_{d}({\it M}^{diag}_{d})^{2} 
  V^{(L)\dagger}_{d}\simeq V_{CKM}({\it M}^{diag}_{d})^{2} 
   V^{\dagger}_{CKM} , 
\label{eq:d-mass2}
\end{equation}
 We note
${\it M}_{D}{\it M}^{\dagger}_{D}$  and ${\it M}_{U}{\it M}^{\dagger}_{U}$ 
 have exactly the same form in the lowest order in $\lambda$.\par
 Next we will determine the form of $M_{U}$ and $M_{D}$.   We define

\begin{equation}
{\it M}_{D} =
\left(
\begin{array}{lcr}
\lambda^{4} & \lambda^{x} & \lambda^{y} \\
\lambda^{a} & \lambda^{2} & \lambda^{z} \\
\lambda^{b} & \lambda^{c} & 1
\end{array}
\right)m_{b} 
\end{equation}
and plug
it into ${\it M}_{D}{\it M}_{D}^{\dagger}$.  Comparing it with
Eq.\  (\ref{eq:d-mass2}), we obtain

\begin{equation}
   x \geq 3, y=3, z=2,  
\end{equation}

\begin{center}
$a,b,c$  are arbitrary.
\end{center}
Similarly we define

\begin{equation}
{\it M}_{U} =
\left(
\begin{array}{lcc}
\lambda^{8} & \lambda^{x} & \lambda^{y} \\
\lambda^{a} & \lambda^{4} & \lambda^{z} \\
\lambda^{b} & \lambda^{c} & 1
\end{array}
\right)m_{t}
\end{equation}
and substitute it into ${\it M}_{U}{\it M}^{\dagger}_{U}$.  Comparing it with 
Eq.(\ref{eq:u-mass2}) we get
\begin{equation}
   x \geq 3, y=3, z=2 
\end{equation}
 \begin{center}
$a,b,c$  are arbitrary.
\end{center}
  If we put $\lambda^{7}$ instead of $\lambda^{8}$ in
the  (1,1) component, we obtain the same result.\par
  The superpotential of the down quarks is given by

\begin{equation}
   W \sim \lambda^{(d)}_{ij}Q_{i}d^{(c)}_{j}H_{1}\left( \frac{\theta}{M_{P}}
 \right)^{n^{(d)}_{ij}} .  
\end{equation}
From the conservation of chiral $U(1)_{X}$  charge we get \cite{E1}
\begin{eqnarray}
    n^{(d)}_{ij}+n^{(d)}_{ji} &=&  n^{(d)}_{ii}+n^{(d)}_{jj}  \\
    & =& \alpha_{i}+\alpha_{j}+\gamma_{i}+\gamma_{j}-2(\alpha_{3}+\gamma_{3})
   \nonumber  
\end{eqnarray}
and using this we can get the lower components  $ n^{(d)}_{ij} (i>j)$.
 
Similar relations hold for the up quarks.  The lower triangular components
 of the mass matrix are determined using these relations, and we obtain
\begin{equation}
{\it M}_{U}=
\left(
\begin{array}{lcc}
\lambda^{8}(\lambda^{7}) & \lambda^{x_{u}} & \lambda^{3} \\
\lambda^{12-x_{u}} & \lambda^{4} & \lambda^{2} \\
\lambda^{5} & \lambda^{2} & 1
\end{array}
\right)m_{t} ,
\end{equation}

\begin{equation}
{\it M}_{D}=\lambda^{x}
\left(
\begin{array}{lcc}
\lambda^{4} & \lambda^{x_{d}} & \lambda^{3} \\
\lambda^{6-x_{d}} & \lambda^{2} & \lambda^{2} \\
\lambda & 1 & 1
\end{array}
\right)m_{t} .
\end{equation}
where
\begin{equation}
  \lambda^{x}=\frac{m_{b}}{m_{t}}  \label{xdef}
\end{equation}
The (2,1) components and the prefactor of the matrix $M_{D}$  are not
 determined yet.  In order to satisfy Eq.(\ref{n33}), since the charge of the
 matter fields are non-negative,  $  X(H_{2})\equiv h_{2}\leq 0$.
 Without loss of generality we can assume
$ h_{2}=0$.  Then we get
\begin{equation}
  X(Q_{3})=X(u^{c}_{3})=0 . \label{q3}
\end{equation}
 Moreover we obtain
 $ x_{u}=5 $
 due to conservation of $U(1)_{X}$ charge.  Furthermore,
 the (1,1) component of $M_{U}$ must be $\lambda^{8}$, because in the case of
 $\lambda^{7}$,  $U(1)_{X}$ charge is not conserved. 
 ${\it M}_{U}$  is then determined uniquely as 

\begin{equation}
{\it M}_{U}=
\left(
\begin{array}{lcr}
\lambda^{8} & \lambda^{5} & \lambda^{3} \\
\lambda^{7} & \lambda^{4} & \lambda^{2} \\
\lambda^{5} & \lambda^{2} & 1
\end{array}
\right)m_{t} ,
\end{equation}
and we can determine the $U(1)_{X}$ charges of up quarks, as follows,
\begin{equation}
 X(Q_{2})=2, X(Q_{1})=3, X(u_{2}^{c})=2, X(u^{c}_{1})=5 .  \label{uch}
\end{equation}

Once $U(1)_{X}$  charges of up quarks are fixed, $x_{d}$ is determined
as $x_{d}=3$.   Then we get

\begin{equation}
{\it M}_{D}=\lambda^{x}
\left(
\begin{array}{lcc}
\lambda^{4} & \lambda^{3} & \lambda^{3} \\
\lambda^{3} & \lambda^{2} & \lambda^{2} \\
\lambda & 1 & 1
\end{array}
\right)m_{t} ,
\end{equation}
and $U(1)_{X}$ charges of down quarks are given by
\begin{eqnarray*}
 X(d_{3L}) &=& 0,\  X(d_{2L})=2,  \\ X(d_{1L}) &=& 3,\   
 X(d^{c}_{3})=x-h_{1} \equiv \gamma_{3},               \label{dch}
\end{eqnarray*}
\begin{equation}
  X(d^{c}_{2})=x-h_{1}=\gamma_{3},\  X(d^{c}_{1})=x-h_{1}+1=\gamma_{3}+1 . 
\end{equation}
In this way quark mass matrices are determined uniquely.  From the mass
 matrices determined this way, the $U(1)_{X}$ charges of quarks are determined
 excepting $x$ and $h_{1}$ as in Eqs.(\ref{q3}), (\ref{uch}) and (\ref{dch}).\\

\bigskip
\textbf{Higgs Sector}\par
There remains undetermined the $U(1)_{X}$ charge of $H_{1}$, i.e., $h_{1}$.
  In the low energy effective superpotential, the term
\begin{equation}
   W=\mu H_{1}H_{2}  
\end{equation}
is needed to avoid the (electroweak scale) axion and to get Higgsino mass.
Phenomenologically, $\mu \sim m_{Z}$ , where $m_{Z}$ is the mass of the
 neutral weak gauge boson.  According to string theory,
 at least three fields are multiplied in the superpotential.  Then $\mu$ must
 be a kind of vacuum expectation value of some field(s) which might be a 
(standard model) gauge singlet or a condensate or a product of several
 fields.  We denote it as ``N''.  Then $\langle N\rangle =\mu$ and
\begin{equation} 
  W=\langle N\rangle H_{1}H_{2}  
\end{equation} 
In the case of $N \sim \theta^{n}$, i.e., 
\begin{equation}
   W=\lambda_{H}H_{1}H_{2}\frac{\theta^{n}}{M^{n-1}_{P}}, \label{hmu}
\end{equation}
the natural scale of $\mu$ would be the string scale.  So we have to forbid
 this term.  Since $h_{2}=0$, we get $h_{1}\leq 0.$  Then from
 Eq.(\ref{hmu}),
 $X(N)\geq 0$ in general.  When $h_{1}<0$, $m^{2}_{H_{1}}$ might be negative
at the string scale, and then the vacuum becomes unstable.  Even if it is not
negative, because $m^{2}_{H_{1}} \ll m^{2}_{H_{2}}$ at the string scale,
and $\tan\beta \equiv \langle H^{0}_{2}\rangle/\langle H^{0}_{1}\rangle$
 might become 
much smaller than one, it is phenomenologically unacceptable. 
 Furthermore, because of naturalness of the electroweak scale,  mass of Higgses
should be $\lsim $ 1TeV \cite{N1}.  If $h_{1}=0$ , it can be
 satisfied.  Hence we will assign 
$ h_{1}=0$ .
Namely two (electroweak) Higgs supermultiplets have vanishing $U(1)_{X}$
 charge.  Then $X(N)=0$.  When ``N'' is in the observable sector (wall),
the vacuum expectation value of which is generally string scale, $\mu$
 is too large.  When
``N'' is in the hidden sector (another wall), there is no direct coupling
to Higgses.  It has to be mediated by fields in the bulk.  Then the $\mu$-term
 may become of electroweak scale, but the scenario is much model-dependent, 
and there is a shade of fine-tuning.  We therefore
assume as a more natural scheme that the $\mu$-term is generated through
 K\"{a}hler  potential by the
Giudice-Masiero mechanism\cite{G2}.  There is a model \cite{A1}, for example,
 in which
\begin{equation}
   K(T,U)+K_{Higgs} \sim -{\rm ln}[(T+\bar{T})(U+\bar{U})-(H_{1}+\bar{H}_{2})
 (\bar{H}_{1}+H_{2})] .  
\end{equation}
where
\begin{equation}
  K=K(S)+K(\Phi)+K_{Higgs}+K_{matter}(Q,Q^{\dagger})+K(\theta,\theta^{\dagger})
 \label{Kahler}
\end{equation}
in which $S$ is dilaton and $\Phi$ denotes moduli $T_{i}$ or $U_{m}$.
The effective superpotential is then
\begin{equation}
   W\sim m_{\frac{3}{2}}H_{1}H_{2} .   \label{mu}
\end{equation}  
Thus we have determined the $U(1)_{X}$ charges of Higgses and also obtained the
 $\mu$-term of the electroweak scale consistently.

\bigskip
{\bf The sparticle spectrum}  \par
 We assume (scalar) particles which have nonzero $U(1)_{X}$ charges are
 in flat directions, otherwise they get mass of the Planck scale.
   To be concrete, they are those particles of the
 standard model and $\theta$-field.  Hence they are massless at high energy.
The $D$-term contribution to the scalar potential is
\begin{equation}
    V_{D}=\frac{1}{2}g^{2}_{X}D^{2}_{X}  
\end{equation}
where  
\begin{equation}
   D_{X}=\sum_{i}K_{i\bar{j}}\phi^{\ast}_{\bar{j}}X\phi_{i}-\theta^
{\ast}\theta+\xi^{2}+\cdots
\end{equation}
and $K_{i\bar{j}}$ denotes $\partial^{2}K/\partial\phi_{i}\partial
 {\bar{\phi}}_
{\bar{j}}$.
A Fayet-Iliopoulos term is generated at the one string loop level from the
anomaly cancellation by the Green-Schwarz mechanism
 and is given by \cite{A2}

\begin{equation}
  \xi^{2}=\frac{{\rm Tr}X}{192\pi^{2}}M^{2}_{P} .
\end{equation}  
Since ${\rm Tr} X\neq 0$ as a result of the $U(1)$ anomaly, $<\theta>$ is 
nonzero and the value of which is 
 near the string scale\footnote{In M-theory the
scale of $M_{string}$, $M_{GUT}$, or $M_{pl}$ should be sometimes replaced
with five-dimensional Planck scale, five-dimensional radius, etc., but
 the change of scale is rather small.} in order to be $D_{X}=0$.  The
 $U(1)_{X}$ 
symmetry is broken but SUSY is not broken yet.  \par

  We assume SUSY breaking occurs through gaugino condensation in the
 hidden sector, and it is communicated to the observable sector
 gravitationally.  Gravitino mass is then given by
\begin{equation}
  m_{\frac{3}{2}}=e^{\frac{1}{2}<K>}|W| .
\end{equation}
When the cosmological constant vanishes, it is given by 
\begin{equation}
  m^{2}_{\frac{3}{2}}= \frac{1}{3} \langle K_{i\bar{j}}
 F^{i}\bar{F}^{\bar{j}}\rangle .  
\end{equation}
where $F^{i}$ is an auxiliary field of a chiral field $\phi_{i}$. 
 Especially if
 SUSY breaking is caused by dilaton and moduli because either $F^{S}$ or
 $F^{T}$ gets nonzero value owing to gaugino condensation,
 $\langle \lambda\lambda\rangle \neq 0$, and if 
\begin{equation}
   K(S)+K(\Phi)=-{\rm ln}(S+\bar{S})-3{\rm ln}(T+\bar{T}) ,  
\end{equation}
    at the tree level as in string theory and M-theory in Eq.(\ref{Kahler}),
 then
\begin{equation}
  m^{2}_{\frac{3}{2}}=\frac{|F^{S}|^{2}}{3(S+\bar{S})^{2}}
 +\frac{|F^{T}|^{2}}{(T+\bar{T})^{2}} \sim \frac{\langle \lambda\lambda
 \rangle^{2}}{M^{4}_{P}} . 
\end{equation}  \par
However, the magnitude or scale of $F^{S}$ or $F^{T}$ is not definite. 
 They depend on various factors: 
 the group structure of the hidden sector, the existence of hidden matter,
 gaugino condensation scale,
 the gauge coupling function, the structure of moduli space, the existence
 of five-branes and so on. 
 In M-theory it is said that $F^{S} \sim F^{T}$\cite{N2}.  The value of
 $\langle S \rangle$  or $\langle T\rangle$  is not known yet.  We assume here
 $m_{\frac{3}{2}}$ is around 1TeV.  In other words, $F^{S}$ and/or $F^{T}$ 
should be of order $10^{11}-10^{13}$ GeV.  Because fifth dimension $\rho$
 is near $\rho_{crit}$ \cite{W1} , the gauge coupling constant in the hidden 
sector is rather
 large in the case of $E_{8}$ and the condensation scale may become
 too large.  The non-standard
gauge-embedding would be needed and the gauge group in the hidden sector
would have to be broken to a smaller one in order to have the condensation
scale of near $10^{11}-10^{13}$ GeV.    
    \par
Soft scalar masses are
 written as 
\begin{equation}
  m_{I\bar{J}}^{2}=m_{I\bar{J}}^{2}|_{F}+m_{I\bar{J}}^{2}|_{D}
 \label{softmass} ,
\end{equation}
where \cite{K2}
\begin{equation}
   m^{2}_{I\bar{J}}|_{F}=Z_{I\bar{J}}m^{2}_{\frac{3}{2}}-F^{i}\bar{F}
 ^{\bar{j}}[\partial_{i}\bar{\partial}_{\bar{j}}Z_{I\bar{J}}-Z^{L\bar{N}}
 \partial_{i}Z_{I\bar{N}} \bar{\partial}_{\bar{j}}Z_{L\bar{J}}]. 
                                     \label{softF}    
\end{equation}
and
\begin{equation}
 K_{matter}(Q,\bar{Q})=Z_{\bar{I}J}(\Phi,\bar{\Phi})\bar{Q}^
{\bar{I}}Q^{J}+\cdots.
\end{equation}
Subscript $I$ denotes matter and $i$ denotes dilaton and moduli.  Soft terms
 derived from
 Eq.(\ref{softF}) is model-dependent much.  They depend on the modular weight 
of each quark, and on whether $F^{S}$ or $F^{T}$ is larger than the other, 
and on the form of the K\"{a}hler potential.  Similarly scalar masses in 
Eq.(\ref{softF}) at
 low energy
 are affected strongly by the renormalization group and very much
 model-dependent.  So even if two kinds of squarks have an identical mass, 
they are different at low energy in general.   The mass difference of squarks
 as a result of Eq.(\ref{softF}) would cause 
the problem of the flavor-changing neutral current and many parameters would 
cause CP breaking.  To avoid these problems it would be necessary to have 
very precise universality of squark masses or alignment with quark matrices.  
These seem to be rather fine-tuning if there is not a certain symmetry which 
requires them.   \par  
However, in our scheme, concurrent with the gaugino condensation the
 value of the $D$--term of the anomalous $U(1)$ symmetry becomes nonzero due
 to the existence of hidden matter, and the $D$-term contribution is much 
greater than the $F$-term contribution.  Consequently the complicated
 situations or various parameters concerning $F$-terms do not matter.  \par    
As discussed before, the gauge group in the hidden 
sector would be broken to the subgroup smaller than $E_{8}$, say $SU(N)$,
 and hidden
 matter would also exist.  
Then the discussion of Binetruy and Dudas 
in \cite{D1} for the case of $SU(N)\times U(1)_{X}$ would be applicable 
also to this case here \footnote{see also ref.\cite{B11}.  It also uses the
 result of Binetruy and Dudas, considering both the $F$-term and $D$-term 
breaking.}.  
The $D$-term becomes nonzero concurrently with gaugino condensation,
 and the position
 of the minimum of the scalar potential is shifted from $\langle \theta 
\rangle \simeq \xi$.  Whereupon the squarks get
 soft terms of 
the following form,
\begin{equation}
 m^{2}_{i}|_{D}=X(\phi_{i})g^{2}_{X}\langle D_{X}\rangle.  \label{Dmass}
\end{equation}
where the $\langle D_{X} \rangle$ is given by 
\begin{equation}
\langle D_{X}\rangle \sim \left[\frac{N_{f}\langle \lambda\lambda \rangle}
{\xi^{2}}\right]^{2}
\end{equation}
and $N_{f}$ denotes the number of flavors of hidden matter
 ``quarks'' \footnote{The 
characteristic scale of the anomalous $U(1)$ symmetry is $\xi$.  Then also
 from dimensional analysis $m_{i}|_{D}$ would be of the order 
$\langle\lambda\lambda \rangle /\xi^{2}$.}.  Then  
\begin{equation}
 \frac{m_{\frac{3}{2}}}{m_{i}|_{D}} \simeq \frac{\langle \lambda\lambda
 \rangle /M^{2}_{P}}{\langle \lambda\lambda \rangle /\xi^{2}} =
 \left(
  \frac{\xi}{M_{P}} \right)^{2} \simeq \varepsilon^{2}   
\end{equation}
Namely $m^{2}_{i\bar{j}}|_{D}$ due to the anomalous $U(1)_{X}$ contributions 
 are much
 larger than the supergravity-induced soft terms, $m^{2}_{i\bar{j}}|_{F}$
\footnote{In the case of SUSY breaking by the $D$-term,
 $m_{\frac{3}{2}}/m_{i}|_{D} \sim \varepsilon$.}.
Using the Eqs.(\ref{q3})  and (\ref{uch}), which were obtained from the 
discussion of the quark mass,
\begin{equation}
X(t)=X(t^{c})=0, X(c)=X(c^{c})=2, X(u)=3, X(u^{c})=5,
\end{equation}
the squark masses are obtained as the following :
\begin{equation}
m^{2}_{\tilde{t},\tilde{t}^{c}}\simeq m^{2}_{\frac{3}{2}},
m^{2}_{\tilde{c},\tilde{c}^{c}}\simeq m^{2}_{\frac{3}{2}}
(1+\frac{2}{\varepsilon^{4}}),
m^{2}_{\tilde{u}}\simeq m^{2}_{\frac{3}{2}}(1+\frac{3}{\varepsilon^{4}}),
m^{2}_{\tilde{u}^{c}}\simeq m^{2}_{\frac{3}{2}}(1+\frac{5}{\varepsilon^{4}}).
\end{equation}
Similarly, since $h_{1}=0$ , we could determine the $U(1)_{X}$ charges of down 
quarks as follows :
\begin{eqnarray*}
  X(b)=0, X(b^{c})=x, X(s)=2, \\
\end{eqnarray*}
\begin{equation}
  X(s^{c})=x, X(d)=3, X(d^{c})=x+1.
\end{equation}
where $x$ is defined in Eq.(\ref{xdef}).  We obtain
\begin{eqnarray*}
m^{2}_{\tilde{b}}\simeq m^{2}_{\frac{3}{2}},
m^{2}_{\tilde{b}^{c}}\simeq m^{2}_{\frac{3}{2}}(1+\frac{x}{\varepsilon^{4}}),
m^{2}_{\tilde{s}}\simeq m^{2}_{\frac{3}{2}}(1+\frac{2}{\varepsilon^{4}}),
\end{eqnarray*}
\begin{equation}
m^{2}_{\tilde{s}^{c}}\simeq m^{2}_{\frac{3}{2}}(1+\frac{x}{\varepsilon^{4}}),
m^{2}_{\tilde{d}}\simeq m^{2}_{\frac{3}{2}}(1+\frac{3}{\varepsilon^{4}}),
m^{2}_{\tilde{d}^{c}}\simeq m^{2}_{\frac{3}{2}}(1+\frac{x+1}{\varepsilon^{4}})
 .
\end{equation}
To be consistent with observed quark masses, $1\leq x \leq 3$,  probably $x=2$
 or
 3 is better than the choice $x=1$, because in our scheme
 $\tan\beta=v_{2}/v_{1}
\sim 1$.   \par
From the above result, the mass of top and left-handed bottom squarks would
be around the gravitino mass, i.e., around 1TeV, meanwhile other squarks would
be roughly $10-100$ TeV for $\varepsilon \sim 0.1-0.2$.  Although the
 right-handed
 bottom squark is heavy,
 because
 of the smallness of the factor $\lambda^{x}$ in the Yukawa coupling
 ($\lambda \simeq 0.22$),  
it would not upset the
 naturalness of the electroweak scale \cite{N1}.  
 Therefore we are able to avoid
 too large contribution of the
 flavor-changing neutral current (FCNC) as well \cite{97d11}.  Namely, because
 first and second
generation squarks are heavy enough, their contributions to FCNC processes
are very small \cite{N1}. \footnote{There may also occur problems
 in the above decoupling scenario taking higher-loop (two-loop) corrections
 into
 consideration.  
But the contribution of higher-loop effects is not clear yet.}. \par

As we have only a single chiral $\theta$ field, and not a vectorlike pair,
the gaugino mass is not caused by $<\theta>\neq 0$ at the tree level.
Accordingly the gaugino mass is generated by the F terms gravitationally
 and by loop effects.  Gravitationally, the gaugino mass is given by
\cite{K2},\cite{S1}
\begin{equation}
\tilde{m}_{a}=\frac{F^{\phi}}{2\Re\it{e}(f_{a})}\frac{\partial f_{a}}
{\partial \phi}
\end{equation}
where $f_{a}$  is a gauge kinetic function in the observable sector.  
When $f_{a}$ is expressed as 
\begin{equation}
   f_{a}=S+\alpha T,  
\end{equation}
then
\begin{equation}
   \tilde{m}_{\frac{1}{2}}=\frac{F^{S}+\alpha F^{T}}
   {2\Re\it{e}(S+\alpha T)} .  
\end{equation}
  It is expected to be of the order
of the gravitino mass from M-theory in contrast to other scenarios where
 $m_{\frac{1}{2}}$ is estimated to be a much smaller value which is 
phenomenologically awkward. \par
 In \cite{C1} it is stated that if the
 first two
 families of sparticles are heavier than $\sim 10m_{\tilde{g}}$ (where
 $m_{\tilde{g}}$ is the gluino mass and $m_{\tilde{g}} \sim m_{\frac{3}{2}}$)
 while third family squarks are heavier
 than $\sim 550$GeV, then none of the complex phases of the soft terms
 lead to unacceptable EDMs for the electron or the neutron, even if
 CP violation is maximal.  In the MSSM there are more than 40 physical 
CP violating phases.  To solve the CP problem by the anomalous $U(1)$ symmetry
seems much simpler than to assume ``ad hoc'' symmetries in order to make very
many complex phases real.  Furthermore, in the latter scenario there may be 
subtlety in running down from the string scale to the electroweak scale
by the renormalization group method.  The explanation which was done in this
paper conforms to the conditions mentioned above.  \par     
   $B$-term is given by \cite{K2}
\begin{eqnarray}
  B&=&F^{i}\{\partial_{i}\mu+\frac{1}{2}\mu K_{i}-Z^{N\bar{J}}
 \partial_{i}Z_{\bar{J}(I}\mu_{J)N}\}-m_{\frac{3}{2}}\mu \nonumber \\
& \cong& \frac{1}{2}\mu(K_{S}F^{S}+K_{T_{i}}F^{T_{i}})-m_{\frac{3}{2}}\mu
  -\mu F^{T_{i}}Z^{H_{1}\bar{H}_{2}}(\partial_{T_{i}}Z_{\bar{H}_{2}H_{1}}) .
\end{eqnarray}
In generic models typically $B \gg  \mu^{2}$ where $B \sim \mu^{2}$ is needed
 phenomenologically. Here it is estimated to be of the order
 $m_{\frac{3}{2}}^{2}$  because
 $\mu \sim m_{\frac{3}{2}}$ was obtained as in Eq.(\ref{mu}).

As for the $A$ terms the largest one is $A^{u}_{33}$,  since $\lambda^{u}_{33}
\sim 1$.  To evaluate $A^{u}_{33}$, the value of  $\bar{F}^{\bar{\theta}}$ is  

\begin{equation}
 \bar{F}^{\bar{\theta}}=e^{\frac{1}{2}K}K^{\theta\bar{\theta}}
 (\partial_{\theta}W+W\partial_{\theta}K)
 \cong e^{\frac{1}{2}K}\theta^{\dagger}W
\end{equation}
because $\partial_{\theta}W=0$ by assumption.  Then

\begin{equation}
  A^{u}_{33}  \\
   =F^{\theta}\{\partial_{\theta}(e^{\frac{K}{2}}
    \lambda^{u}_{33})+\frac{1}{2}K_{\theta}e^{\frac{1}{2}K}
  \lambda^{u}_{33}\}  \\
  \sim\varepsilon^{2}m_{\frac{3}{2}}e^{\frac{1}{2}K} ,
\end{equation}
so it satisfies the condition that $A \lsim 3$.  The values
satisfy the constraint on $A$ and $B$.  It also affects the CP violation by
 soft terms to a better direction.

\bigskip
{\bf Discussion}

From the definition of $\varepsilon$, the value of $\varepsilon$ is given by
\begin{equation}
   \varepsilon=\frac{\langle \theta \rangle}{M_{P}} \sim \frac{\xi}{M_{P}}
 \sim 
  \sqrt{\frac{1}{192\pi^{2}}\sum_{i}X(\phi_{i})} \sim 0.1
\end{equation}  
if we suppose the sum of the chiral charge of leptons are the same order 
as that of quarks.
Thus we have obtained $\varepsilon$ to be the same order as the 
Wolfenstein parameter, $\lambda \sim 0.2$.     \par
We did not mention to the mass of fermionic $\theta$.  But its interaction 
with ordinary matter is very weak because the $U(1)_{X}$ gauge boson is
very heavy.  \par
  The problem of the flavor- changing
 neutral current and CP phases may be solved simply using anomalous $U(1)$
 symmetry besides having explained  quark mass hierarchy almost uniquely. 
 Because the $D$-term is nonzero , the cosmological constant seems to be
 nonzero.  
It is a very
 attractive idea that inflation is triggered by the $D$-term of the anomalous
 $U(1)$ symmetry \cite{B1} .  This is expected to occur also around $10^{13}$
GeV or so, namely near the scale of the gaugino condensation.  \par
If there is an effective nonrenormlizable coupling between the gaugino in the
hidden sector and ``gauge-singlet'' right-handed neutrinos such as
 $\lambda\lambda\nu^{c}_{i}\nu^{c}_{i}$, the right-handed neutrinos get mass
of the order $10^{13}$ GeV.  Then mass of the lepton sector fits well, too.  \\
\indent The most promising idea of baryogenesis is that it is generated by
 the Affleck-
 Dine (AD) mechanism \cite{A3}. 
If the texture of mass matrices of quarks and leptons are similar,
$X(\nu^{c}_{3})$ would be zero.  In the early universe anti-sneutrino of
 the third generation might have a vacuum expectation value,
$<\tilde{\nu}^{c}_{3}>\neq 0$, and break lepton number, and then by sphareron
may be converted to baryon asymmetry \cite{C2}.
Casas and Gelmini \cite{C3} state that if the AD field is not charged
 under the inflationary $U(1)$, the AD mechanism works good.  To regard the
 third generation sneutrino $\tilde{\nu}^{c}_{3}$ as the AD field may fit well.

\bigskip
{\bf Acknowledgements}\par
I am indebted much to H. Sato.  I also thank K. Sudoh for discussion.


\begin{thebibliography}{99}
 \bibitem{R1}
See, for example, P. Ramond, ``Mass Hierarchies from Anomalies'',
 hep-ph/9604251 ; G. Dvali and A. Pomarol, ``Anomalous $U(1)$ as a mediator
of Supersymmetry Breaking'', Phys. Rev. Lett. {\bf 77}
(1996) 3728 ; E. Dudas, C. Grojean, S. Pokorski and C. A. Savoy, ``Abelian
Flavour Symmetries in Supersymmetric Models'', Nucl. Phys. {\bf B481}
 (1996) 85,  and references therein.
\bibitem{H1}
P. Ho\v{r}ava and E. Witten, Nucl. Phys. {\bf B460} (1996) 506 ;
ibid. {\bf B475} (1996) 94.
\bibitem{K1}
  H. Kataoka, H. Munakata, H. Sato and S. Tanaka, Phys. Lett.
      \textbf{B342} (1995) 79.
\bibitem{R2}
 There are too many references to cite here without omission.
 For a review, see, for example,
  P. Ramond,''Mass Hierarchies from Anomalies''  hep-ph/9604251 ;
      ``A Model for Fermion Mass Hierarchies and Mixings'' hep-ph/9808489
 and references therein. 
\bibitem{E1}
  J. K. Elwood, N. Irges and P. Ramond, Phys. Rev. Lett. {\bf 81}
 (1998) 5064. 
\bibitem{G1}
  M. Green and J. Schwarz, Phys. Lett. {\bf B149} (1984) 117. 
\bibitem{I1}
  L. E. Ib\'{a}\~{n}ez, Phys. Lett. {\bf B303} (1993) 55
\bibitem{D1}
  G. Dvali and A. Pomarol, , Phys. Rev.Lett. {\bf 77} (1996) 3728 ;
  P. Binetruy and E. Dudas, Phys. Lett. {\bf B389} (1996) 503 ;
  R. N. Mohapatra and A. Riotto, Phys. Rev. {\bf D55} (1997) 4262 ;
  N. Arkani-Hamed, M. Dine and S. P. Martin, Phys. Lett. {\bf B431} (1998) 329.
\bibitem{C1}
  A. G. Cohen, D. B. Kaplan and A. E. Nelson, Phys. Lett. {\bf B388}
 (1996) 588. 
\bibitem{N1}
  A. E. Nelson and D. Wright, Phys. Rev. {\bf D56} (1997) 1598.
\bibitem{F1}
  C. D. Froggatt and H. B. Nielsen, Nucl. Phys.  {\bf B147} (1979) 277 ;
  B. Pendleton and G. G. Ross, Phys. Lett. {\bf B98} (1981) 291 ;
  C. T. Hill, Phys. Rev. {\bf D24} (1981) 691 ;
  A. Bardeen, C. T. Hill and M. Lindner, Phys. Rev. {\bf D41} (1990) 1647.
\bibitem{G2}
  G. Giudice and A. Masiero, Phys. Lett. {\bf B206} (1988) 480.
\bibitem{A1}
  I. Antoniadis, E. Gava, K. S. Narain and T. R. Taylor,
    Nucl. Phys. {\bf B432} (1994) 187.  
\bibitem{A2}
  J. Atick, L. Dixon and A. Sen, Nucl. Phys. {\bf B292} (1987) 109 ; 
  M. Dine, I. Ichinose and N. Seiberg, Nucl. Phys. {\bf B293} (1987) 253.
\bibitem{N2}
  see, for example, H. P. Nilles, M. Olechowski and M. Yamaguchi, Phys.
 Lett. {\bf B415} (1997) 24 ; Z. Lalak and S. Thomas, Nucl. Phys. {\bf B515}
 (1998) 55 ; A. Lukas, B. A. Ovrut and D. Waldram, Phys. Rev. {\bf D57}
 (1998) 7529 ; T. Li, J. L. Lopez and D. V. Nanopulos, Phys. Rev.{\bf D56}
 (1997) 2602. 
\bibitem{W1}
  E.Witten, Nucl. Phys. {\bf B471} (1996) 135.  
\bibitem{K2}
  V. S. Kaplunovsky and J. Louis, Phys. Lett. {\bf B306} (1993) 269 ;
  J. Louis and Y. Nir, Nucl. Phys. {\bf B447} (1995) 18.  
\bibitem{B11}
  P. Binetruy, C. Deffayet, E. Dudas and P. Ramond, Phys. Lett. {\bf B441} (1998) 163. 
\bibitem{97d11}
K. Dienes and C. Kolda,''Twenty Open Questions in Supersymmetric
Particle Physics'', hep-ph/9712322; V. Barger, C.Kao and R.Zhang, hep-ph/9911510.
\bibitem{S1}
  S. K. Soni and H. A. Weldon, Phys. Lett. {\bf B126} (1983) 215. 
\bibitem{B1}
  P. Binetruy and G. Dvali, Phys. Lett. {\bf B388} (1996) 241 ;
 E. Halyo, Phys. Lett. {\bf B387} (1996) 43 ;
 T. Matsuda, Phys. Lett. {\bf B423} (1998) 35
\bibitem{A3}
  I. Affleck and M. Dine, Nucl. Phys. {\bf B249} (1985) 361. 
\bibitem{C2}
  B. A. Campbell, S. Davidson and K. A. Olive, Phys. Lett. {\bf B303}
 (1993) 63 ;
            Nucl. Phys. {\bf B399} (1993) 111. 
\bibitem{C3}
  J. A. Casas and G. B. Gelmini, Phys. Lett. {\bf B410} (1997) 36. 
\end{thebibliography}
\end{document}